\documentstyle[12pt]{report}
\begin{document}
\oddsidemargin=0in
\textwidth=6.25in
\topmargin=0in
\textheight=609pt
\parskip=14pt
\setlength{\unitlength}{0.5cm}
\def\adj{\mathop{\rm adj}}
\def\diag{\mathop{\rm diag}}
\def\rank{\mathop{\rm rank}}
\def\span{\mathop{\rm span}}

\def\rdots{\mathinner{\mkern1mu\raise1pt\vbox{\kern1pt\hbox{.}}\mkern2mu
   \raise4pt\hbox{.}\mkern2mu\raise7pt\hbox{.}\mkern1mu}}
\newcommand{\Z}{{\rm Z\kern-.35em Z}}
\newcommand{\bP}{{\rm I\kern-.15em P}}
\newcommand{\Q}{\kern.3em\rule{.07em}{.65em}\kern-.3em{\rm Q}}
\newcommand{\R}{{\rm I\kern-.15em R}}
\newcommand{\h}{{\rm I\kern-.15em H}}
\newcommand{\C}{\kern.3em\rule{.07em}{.55em}\kern-.3em{\rm C}}
\newcommand{\T}{{\rm T\kern-.35em T}}
\newcommand{\ojo}{\marginpar{$\Leftarrow$}}
\newtheorem{theorem}{Theorem}
\newtheorem{example}{Example}
\newtheorem{rem}{Remark}
\newtheorem{lem}{Lemma}
\newtheorem{proposition}{Proposition}
\newtheorem{corollary}{Corollary}
\newtheorem{definition}{Definition}

\vfil
\eject

\centerline{A New Formulation and Regularization of Gauge Theories}
\centerline{Using a Non-Linear Wavelet Expansion$^*$}

\vspace{1.0in}

\centerline{Paul Federbush}
\centerline{Department of Mathematics}
\centerline{University of Michigan}
\centerline{Ann Arbor, MI 48109-1003}
\centerline{(pfed@math.lsa.umich.edu)}
\vspace{4.0in}

$^*$ This work was supported in part by the National Science Foundation
under Grant No. PHY-92-04824.

\vfill\eject

\centerline{ABSTRACT}

\indent
The Euclidean version of the Yang-Mills theory is studied in four
dimensions.  The field is expressed non-linearly in terms of the basic
variables.  The field is developed inductively, adding one excitation at a
time.  A given excitation is added into the ``background field'' of the
excitations already added, the background field expressed in a radially
axial gauge about the point where the excitation is centered.  The
linearization of the resultant expression for the field is an expansion
$$
A_\mu(x) \ \cong \  \sum_\alpha \; c_\alpha \; \psi_\mu^\alpha(x)
$$
where $\psi^\alpha_\mu(x)$ is a divergence-free wavelet and $c_\alpha$ is
the associated basic variable, a Lie Algebra element of the gauge group.
One is working in a particular gauge, regularization is simply cutoff
regularization realized by omitting wavelet excitations below a certain
length scale.  We will prove in a later paper that only the usual
gauge-invariant counterterms are required to renormalize perturbation theory.

Using related ideas, but essentially independent of the rest of paper, we
find an expression for the determinant of a gauged boson or fermion field
in a fixed ``small'' external gauge field.  This determinant is expressed
in terms of explicitly gauge invariant quantities, and again may be
regularized by a cutoff regularization.

We leave to later work relating these regularizations to the usual
dimensional regularization.
\vfill\eject

\chapter{Introduction}
It is the immediate goal of this paper to present some ideas developed in
Constructive Quantum Field Theory, and indicate how we hope to use them to
develop an alternate regularization of Yang-Mills perturbation theory.  It
is our hope that these ideas will find further application in particle
theory.  Our major long range goal only brushed on in this paper is to
adapt the same approach to make contributions to the theory of Quantum
Gravity.

Constructive Quantum Field Theory as pioneered by Glimm and Jaffe has been
completely successful in treating superrenormalizable field theories.
$\phi^4_3$ was the crucial test case, first handled by Glimm and Jaffe in
an impossibly difficult paper [1], and thereafter by many people with much
simplification.  The program we follow was begun in [2] with a treatment of
$\phi^4_3$.  A rather beautiful easy to read exposition of Constructive
Quantum Field theory at this level appears in [3].

Several groups of researchers have worked to go beyond superrenormalizable
field theories, with some success.  We particularly note studies of four
dimensional Yang-Mills theories [4], [5], [6].  There is the general belief
that one already has the mathematical sophistication to prove the rigorous
existence of asymptotically free field theories in the
Euclidean domain ---but that this task is so difficult and complex that no
readable and accepted proof will soon be extant.  We aim for less, for
contributions at the level of perturbation theory or slightly beyond.

Wavelets are discussed in Chapter 2.  The first main idea that is advocated
in this paper is to EXPAND THE FIELD IN TERMS OF WAVELETS (instead of
Fourier components).  For a scalar field $\phi(x)$ ,as in the $\phi^4_3$
theory, such an expansion would be
\begin{equation}
\phi(x) \ \ = \ \ \sum_\alpha \; c_\alpha \; u_\alpha(x).
\end{equation}
The $u_\alpha$ are the wavelets and the $c_\alpha$ are the expansion
coefficients.

The functional integral defining the field theory would be
\begin{equation}
\big< \, {\mit p} \, \big> \ = \ {\cal N} \ \prod_\alpha \left(
\int^\infty_{-\infty} \; dc_\alpha \right) e^{-S} \; {\mit p}
\end{equation}
where $S$ is the action
\begin{equation}
S = \int \left[ {1 \over 2} (\nabla \phi)^2 + {1 \over 2} \; m^2\phi^2 +
\lambda \phi^4 \right]
\end{equation}
expressed in terms of the variables $c_\alpha$.  Making precise sense of
this procedure is the result of [2], [3].

Modern proofs of the renormalizability of $\phi^4_4$ have been given using
ideas of joint localization in $x$-space and $p$-space [7].  This was a
development originating in the renormalization group of K. Wilson and L.
Kadanoff.  Completely parallel proofs using wavelets must be easy to
present; but no one has yet written this down.  We intend to extend results
such as in [7] to the Yang-Mills theory by the use of wavelets.

A perturbation theory developed in terms of wavelets (for $\phi^4_4$ or
Yang-Mills) will not be computationally effective.  It will not compete
with the dimensional regularization of `t Hooft and Veltman [8].
Expressions will be given as infinite sums of complicated integrals
involving wavelets; the expressions will be only valid in Euclidean space,
and cannot be analytically continued term by term.  The expressions in this
paper for the determinant expansions give one an idea what to expect.  But
there should be some theoretical applications.  For quantum gravity where
usual perturbation theory fails there is more flexibility to go beyond
perturbation theory in our formalism, than using dimensional regularization.

We come to the second main idea we use in our approach to Gauge theories.
Unlike the use of wavelets it has not yet had the test of time to prove it
is useful and important.  We call the construction we now describe ``gauge
invariant coupling'' (presented for the lattice Yang-Mills theory in [4],
and for the non-linear sigma model in [9]).  EXCITATIONS ARE ADDED INTO THE
FIELD ONE AT A TIME, in a length scale decreasing order.  BEFORE THE nth
EXCITATION IS ADDED THE FIELD AS SO FAR CONSTRUCTED CONTAINING $n-1$
EXCITATIONS IS FLATTENED OUT AS MUCH AS POSSIBLE NEAR THE EXCITATION BY A
GAUGE TRANSFORMATION.  After the excitation is added one undoes the same
gauge transformation (applies the inverse gauge transformation).

1) \ \ \ For a scalar field $\phi$ in the $\phi^4_3$ theory, there is no
gauge group, and this construction is the usual addition of excitations
given in (1.1).

2) \ \ \ For the non-linear sigma model the gauge group consists of global
gauge transformations.  The construction is given in [9] for certain sigma
models.

3) \ \ \ For the Yang-Mills theory treated herein the gauge group describes
local gauge transformations.  Flattening out the background field near a
point is done by going to a radial axial gauge about that point.  Detailed
description of this case follows below in the discussion embracing
equations (3.8) and (3.9) and the explanation following.

4) \ \ \ In quantum gravity the gauge group is the group of
diffeomorphisms.  Flattening out the background metric of excitations already
added near a point corresponds to passing to a Riemannian normal coordinate
system.

The construction of the theory in terms of our basic variables is complex.
The field is a non-linear function of the variables.  The action is an
infinite degree polynomial in these variables.  But regularization can be
done by cutoff, discarding excitations below some length scale.  Chapter 3
presents the basic formalism for the pure Yang-Mills theory.

Determinants for gauged boson or fermion particles in an external gauge
field are developed in Chapters 4 and 5.  These constructions are not
identical to the treatment of boson or fermion fields coupled to a
Yang-Mills field (one that is not external, but interacting).  We present
the construction of the determinants as interesting in its own right, and
as giving insight into the full theory.  The proofs of the convergence of
the cutoff determinants (renormalized by the usual $c \int |F|^2$
counterterm) as the cutoffs are removed will be much easier than for the
full theory, and the energetic reader will anticipate the ineluctable lines of
these
proofs.  In all cases we will want to relate cutoff limits to the results
of other regularizations.
\vfill\eject

\chapter{WAVELETS}

\indent

For us a set of wavelets, $f_\alpha(x)$, on $R^d$ are a set of functions
orthonormal in some norm, given as translations and scalings of a finite
set of functions.  More explicitly, there are a set of ``mother wavelets'',
$F_t(x)$, $t$ in a finite set $I$, such that the set of functions
\begin{equation}
2^{(d/2)r} \; F_t(2^rx - n) = f_{(r,\gamma,t)}(x) = f_\alpha(x)
\end{equation}
with $r \in Z$, $n \in Z^d, \ t \in I$ are the orthonormal set.  Here
$\alpha$ indicates a specification of $(r, \gamma, t)$
\begin{equation}
\alpha \longleftrightarrow (r,\gamma,t) = \Big( r(\alpha), \gamma(\alpha),
t(\alpha) \Big)
\end{equation}
and $\gamma$ is given as
\begin{equation}
\gamma \ \ = \ \ 2^{-r} \ (n + a_0), \ a_0 = \left( {1 \over 2}, {1 \over
2}, \cdots, {1 \over 2} \right)
\end{equation}
and is thus a point in $R^d$.  $f_\alpha(x)$ is ``centered'' about
$\gamma(\alpha)$ (we also write this as $\gamma_\alpha)$.  $f_\alpha(x)$
may be viewed as ``living'' approximately in a cube of side length $2^{-r}
= \ell_r$ centered at $\gamma_\alpha$.  $\ell_r$ is also called the
``length scale'' of the wavelet.

The wavelets we will use satisfy the following additional properties:
\begin{description}
\item [a)] The functions are ${\cal C}^\infty$, infinitely differentiable.
\item [b)] The functions fall off at infinity faster than any power.
\item [c)] All moments equal zero.  That is
\begin{equation}
\int^\infty_{-\infty} dx_1 \; x^{a_1}_1 \cdots
\int^\infty_{-\infty} dx_d \; x^{a_d}_d \ f_\alpha(x) = 0
\end{equation}
for all non-negative integers $a_i$.
\end{description}

Scalar wavelets with all the above properties (and complete in $L^2(R^d)$)
were developed by Y. Meyer [10].  (The discovery of these wavelets
initiated the vast activity in the wavelet area.)  For each $s$ the wavelet
set, $\psi^s_\alpha$, used in Chapter 4 below may be expressed in terms of the
Y.
Meyer wavelets, $\hat{\psi}_\alpha$, by the expression:
\begin{equation}
\psi^s_\alpha(x) \ = \left( {1 \over  \Big(-\Delta + m^2\Big)^{s/2}} \
\hat{\psi}_\alpha \right) (x).
\end{equation}

Divergence-free vector wavelets are easy to construct in dimensions two,
four, and eight [11].  The four dimensional wavelets, $\psi_\mu^\alpha$ ,
used in this paper satisfy
\begin{equation}
\partial_\mu \ \psi_\mu^\alpha \ \ =  \ \ 0
\end{equation}
and are a complete set for divergence-free vector fields in the inner
product
\begin{equation}
\Big< \psi^\alpha,\psi^\beta \Big> \  = \  \int d^4 x \  \psi^\alpha_\mu(x)
\ \psi^\beta_\mu(x).
\end{equation}

It is often technically convenient to have a maximum length scale for the
wavelets, by omitting wavelets with $r < 0$ say.  This is particularly true
if there is a non-zero mass scale.  One can complete to an orthonormal
basis by adding extra functions at the $r=0$ scale, [3], [4].  We do not
use such truncation in this paper for simplicity in exposition.
\vfill\eject
\chapter{The Basic Formalism}
We order the set of wavelets, $\{ \psi^\alpha_\mu \}$, in a length scale
$(\ell_r)$ decreasing order.  The ordering is then a mapping ${\cal O}$
from the positive integers to the set of $\alpha$'s labelling the
excitations
\begin{equation}
{\cal O} : Z^+ \longrightarrow \{\alpha\}
\end{equation}
or
\begin{equation}
{\cal O} : i \longrightarrow \alpha(i)
\end{equation}
such that if $i > j$ then $r\left(\alpha(i)\right) \ge r\left( \alpha(j)
\right)$.  We are going to construct the gauge field $A_\mu(x)$ by adding in
excitations one at a time in the order given by ${\cal O}$ in (3.1).  We
work with a finite set of excitations (via cutoffs) so in the ordering each
excitation has a predecessor and a successor.

We work in a familiar formalism with
\begin{equation}
F_{\mu \nu} = \partial_\mu A_\nu -\partial_\nu A_\mu + [A_\mu, A_\nu]
\end{equation}
and the action of gauge transformation, $u$:
\begin{eqnarray}
A_{\mu} &\longrightarrow& u A_\mu u^{-1} + u\partial_\mu u^{-1} \\
F_{\mu \nu} &\longrightarrow& u F_{\mu \nu}u^{-1}.
\end{eqnarray}
We note that if $A_\mu$ is pure gauge then
\begin{equation}
A_\mu = u \partial_\mu u^{-1}
\end{equation}
with
\begin{equation}
u^{-1} \ = \ T \ e^{\int^x A_\mu (\bar{x})d\bar{x}^\mu}
\end{equation}
with $T$ ``path ordering''.

We start with $A_\mu = 0$ and add in excitations one at a time in the order
given by ${\cal O}$ above.  We define $B^n_\mu(x)$ to be the field
constructed with the first $n-1$ excitations, it is the ``background
field'' into which the $n^{th}$ excitation is added.  We first introduce an
additional construct.  For a given gauge field, $A_\mu$, we define
\begin{equation}
u(x,y) = u(x,y,A) = T\; e^{\int^y_xA_\mu(\bar{x})d\bar{x}^\mu}\ .
\end{equation}
The line integral in the exponent is along the straight line segment
joining $x$ to $y$.  The $T$ indicates a path ordering (similar to the more
familiar time ordering) of the $A_\mu(x)$ (which of course do not commute).
  We now define
\begin{equation}
B^{n+1}_\mu(x) = B^n_\mu(x) + u(x,\gamma_n,B^n)c_n\;u(\gamma_n, x, B^n)
\psi^n_\mu(x).
\end{equation}
We are abbreviating $\alpha(n)$ by $n$ on sub and superscripts of $c,
\gamma,$ and $\psi$.  In a very physically motivated way we are ``adding''
excitation $c_n\psi^n_\mu$ to the background field, $B^n_\mu$, the field as
constructed so far, by
\begin{description}
\item [1)] ``flattening out'' $B^n_\mu$ near the excitation by passing to
the radial axial gauge about $\gamma_n$, the central location of the
excitation.
\item [2)] adding the excitation to this flattened out background field.
\item [3)] undoing the gauge transformation on this sum.
\end{description}
This was the action leading to (3.9).

In (3.9), $c_n$ is an element of the Lie Algebra of the gauge group, the
amplitude of the excitation and one of the basic variables in terms of
which the theory is expressed.  $A_\mu(x)$ the limit as $n$ goes to
infinity of the $B^n_\mu(x)$ is a very non-linear function of these basic
variables, which if linearized would lead to
\begin{equation}
A_\mu(x) \cong \sum c_n \psi_\mu^n(x).
\end{equation}
And in this sense we are working in a gauge that is a ``perturbation'' of
\begin{equation}
{\partial A_\mu \over \partial x^\mu} \ \cong \ 0.
\end{equation}
One result of our definition (3.9) is that if $B^n(x)$ were pure gauge on
the support of $\psi^n(x)$ then $B^{n+1}(x)$ would be a gauge
transformation of $c_n \psi^n$.

The formal functional integral one works with to define the Euclidean
theory is expressed in the following relations:
\begin{equation}
\big< p \big> \ \ = \ \ [p]/[1]
\end{equation}
with
\begin{equation}
[p] = \prod_\alpha \left( \int \; dc_\alpha \right) e^{-{1 \over g^2}S}
\left({\rm Det}\ M \right)^{1 \over 2} \; p
\end{equation}
and
\begin{equation}
S \ = \ \int \; |F|^2 \ .
\end{equation}
$\int dc_\alpha$ is over the Lie Algebra, an integral over $R^s$ with
$s$ the dimension of the Lie Algebra.  Det $M$ is our determinant,
including the Fadeev-Popov factors, and the Jacobian of the change of
variables to the $c_\alpha$, derived together by the same argument as usual
for the F-P determinant.

We proceed to specify $M$.  We let $L_i, i=1,\cdots,s$, be a basis for the
Lie Algebra, and $\{ \psi^\sigma(x)\}$ be the basis of scalar wavelets
(orthonormal in the inner product $\left< \; , \; (-\Delta) \; \right>$) from
which the vector wavelets $\{\psi^\alpha_\mu(x)\}$ are developed by the
construction in [11].  The index $\alpha$ is there associated to a pair
\begin{equation}
\alpha \longleftrightarrow (\sigma, t)
\end{equation}
where $\sigma$ is an index of a scalar wavelet, and $\#\{t\} = 3$.  (The
$t$ are different from the $t$ in Chapter 2.)  The expansion
coefficient $c_\alpha$ is then represented in more detail as
\begin{equation}
c_\alpha \ = \ \sum_i \; c_{\sigma, t, i} \; L_i
\end{equation}
in terms of scalar variables $\{c_{\sigma, t, i}\}$.  We let
$\{b_{\sigma,i}\}$ be scalar variables, coordinates in the infinitesimal
Lie Algebra of local gauge transformations, defined at the origin of the
Lie Algebra, by
\begin{equation}
\left( {\partial \over \partial b_{\sigma,i}} \; uA_\mu u^{-1} \right)(x)
\Big|_0 \ = \big[ A_\mu(x), L_i \big] \; \psi^\sigma(x) +
\partial_\mu\psi^\sigma(x)L_i
\end{equation}
where $u$ represents the gauge transformation (depending on the $b$'s), $u$
set to the identity after the derivative is taken.  The matrix $M$ has
indices labeled by the union of $\{(\sigma,i)\}$ and $\{(\sigma, t, i)\}$.
We define some of the matrix elements:
\bigskip
\begin{equation}
M_{(\sigma,i)(\sigma',i')} \ = \ \sum_\mu \ \int \; d^4x \; Tr \left(
{\partial A_\mu \over \partial b_{\sigma,i}} \  {\partial A_\mu \over \partial
b_{\sigma',i'}} \right)
\end{equation}
\medskip
\begin{equation}
M_{(\sigma,i)(\sigma',t',i')} \ = \ \sum_\mu \ \int \; d^4x \; Tr \left(
{\partial A_\mu \over \partial b_{\sigma,i}} \  {\partial A_\mu \over \partial
c_{\sigma',t',i'}} \right)
\end{equation}
\bigskip
and the other elements similarly.  Derivatives with respect to $b$'s are
interpreted to be as in (3.17), derivatives with respect to $c$'s must be
computed as determined by the complicated construction of the $A_\mu$.
Note that if $L_i$ is selected orthonormal in the trace product, and if all
$c_\alpha$'s are zero, then $M$ is an infinite identity matrix.

\medskip

The key to the mechanism that is at the heart of our organization of
perturbation theory is laid bare in the following physical description.  We
view a perturbation theory ``diagram'' in which each ``line'' corresponds
to a wavelet (labelled by $(\sigma,t)$ if a vector wavelet -- a gluon line,
or by $(\sigma)$ if a scalar wavelet -- a ghost line).  Then {\it if
a ``connected'' subdiagram has all smaller length scale variables than the
lines to which it is connected}, by a simple, Jacobian-equals-one, change
of variables in the subdiagram, {\it the subdiagram may be
represented as coupled to the gauge field outside the subdiagram in a
gauge, for the ``external'' fields, that is close to a radial axial gauge
about some point of the subdiagram}.  Thus the external fields are small
where evaluated.  (A simplified but essentially
accurate picture.)  This unique feature should have some practical
applications, as well as theoretical, and makes one wonder if such an
organization is at all possible with ordinary momentum space Feynman
integrals for gauge theories.  ({\it Aside:}  asked to consider the axial
vector anomaly, I was immediately led by these ideas to the easy
computation $\partial_\mu J^5_\mu(0)$ in the gauge $A_\mu(x) \cong {1 \over
2} \; x_\alpha \; F_{\alpha \mu}(0)$, in configuration space with $m=0$, a
known but somewhat obscure derivation.)

There is a new, physical idea here.
\vfill\eject
\chapter{Determinant of a Gauged Boson Field in an External Gauge Field.}

In this chapter we find an expression for $M = M(A)$ the functional
of the gauge field $A_\mu$ given formally by:
\begin{eqnarray}
M &=& {\rm Det}({\cal B}) \\
{\cal B} &=& (\partial_\mu + A_\mu)^* (\partial_\mu +A_\mu) + m^2
\end{eqnarray}
For each real $s$ we let $\{\psi^s_\alpha\}$ be the scalar wavelet basis
defined in Chapter 2 orthonormal in the following norm:
\begin{equation}
\left< \Big< \psi^s_\alpha, \psi^s_\beta \Big> \right>_s \ = \ \left< \psi^s
_\alpha, \Big( -\Delta + m^2 \Big)^s \   \psi^s_\beta \right>
\end{equation}
$$
 =  \int d^4x \ \psi^s_\alpha(x)  \left( \Big( -\Delta + m^2 \Big)^s \
 \psi^s_\beta \right) (x) = \delta_{\alpha \; \beta}
$$
With $\gamma_\alpha$ the ``center'' of wavelet $\alpha$, we recall the
expression for $u(x,y)$ in (3.8 ) and define
\begin{equation}
\phi^s_\alpha(x) = u(x, \gamma_\alpha) \; \psi^s_\alpha(x).
\end{equation}
We are now set to begin evaluating $M$.

We begin with some formal reductions:
\begin{eqnarray}
M &=& {\rm Det}({\cal B}) \\
&=& {\rm Det}(m_{\alpha \; \beta})
\end{eqnarray}
where
\begin{equation}
m_{\alpha \; \beta} \ \ = \ \ \Big< \psi^1_\alpha, {\cal B} \;
\psi^1_\beta \Big> .
\end{equation}
(We always are interested in $M$ up to a multiplicative constant
independent of $A_\mu$.)  Changing basis we find from (4.6)-(4.7) that:
\begin{equation}
{\rm Det} ({\cal B}) = {\rm Det}(n_{\alpha \; \beta}) /
|{\rm Det}(b_{\alpha \; \beta})|^2
\end{equation}
where
\begin{eqnarray}
n_{\alpha \; \beta} &=& \Big< \phi^1_\alpha, \; {\cal B} \;
\phi^1_\beta \Big> \\
b_{\alpha \; \beta} &=& \Big< \phi^1_\alpha \, , \, \psi^{-1}_\beta \Big>
\end{eqnarray}
We set
\begin{equation}
h_{\alpha \; \beta} \ \ = \ \ n_{\alpha \; \beta} - \delta_{\alpha \; \beta}
\end{equation}
and proceed to separately treat the numerator and denominator determinants
in (4.8).

Starting with the numerator determinant:
\begin{equation}
{\rm Det}(n) = e^{Tr\big(ln(\delta + h)\big)} \ = \ e^{Tr \Big(h -{h^2
\over 2} \cdots \Big)}.
\end{equation}
We look at a single term in this exponential
\begin{equation}
Tr(h^s) = Tr \sum h_{12} \cdot h_{23} \cdots h_{s1}
\end{equation}
where we abbreviate $h_{\alpha_i \, \alpha_j}$ as $h_{ij}$ and the trace
on the right side of (4.13) is over group indices only.  We note that for a
sum over group indices:
\begin{equation}
\sum_i |i \,>< \, i| = \sum_i v\; |i \, >< \, i|v^*
\end{equation}
For any $v$ unitary in group space.  Using this we make the following
replacements in (4.13):
\begin{equation}
h_{ij} \longrightarrow u^{-1} (\gamma_i, \, \gamma_1)
h_{ij} \  u (\gamma_j, \, \gamma_1) = \hat{h}_{ij}
\end{equation}
(We are ``grounding'' all elements to the center of $\alpha_1$.)  We define
for any integer $n$, and points $x_1, \cdots, x_n$:
\begin{equation}
\ell(x_1,x_2, \cdots , x_n) = u(x_1,x_2)u(x_2,x_3) \cdots u(x_n,x_1)
\end{equation}
(an ordered line integral around a loop).  We find
\begin{equation}
\hat{h}_{ij} = \ell^{-1} (\gamma_1,x,\gamma_i) \left[ \left<  \psi_i(x),
\big( u^{-1} \; {\cal B} \; u \big) \ \psi_j(y) \right> - \delta
\right] \ell(\gamma_1,y,\gamma_j)
\end{equation}
Here $\delta$ is the spatial delta function $\delta(x,y)$, and the
Kronecker delta in group indices.  $x$ and $y$ are understood as being
integrated over.  Inside the matrix element, $(u^{-1}\, {\cal B} \,
u)$ is the integral operator (which we have written in an abbreviated
notation)
\begin{equation}
\left( u^{-1} \, {\cal B} \, u \right) =  u^{-1} (x,\gamma_1){\cal
B} \; u(y,\gamma_1).
\end{equation}
To study $u^{-1} \, {\cal B}\,u$ we use the following identity (obtained by
a relatively hard computation)
\begin{equation}
(\partial_\mu + A_\mu)\; u(x,p) \ = \ u(x,p) \partial_\mu \ + \ {\cal E}_\mu
\end{equation}
where
\begin{equation}
{\cal E}_\mu = \int^1_0 dssu(x,p+s(x-p))F_{\alpha
\mu}(p+s(x-p))u(p+s(x-p),x)\; u(x,p)(x-p)^\alpha
\end{equation}
\begin{eqnarray}
\left( u^{-1} \, {\cal B} \, u \right) &=& B_0 \ + \ E \\
B_0 &=& -\Delta + m^2 \\
E &=& (u\partial_\mu)^* \; {\cal E}_\mu \ + \ {\cal E}^*_\mu
(u\partial_\mu) \ + \ {\cal E}^*_\mu \, {\cal E}_\mu
\end{eqnarray}
$E$ is a first order differential operator (so of form $a_\mu(x)\;
\partial_\mu + b(x)\ )$.

Combining (4.13)-(4.17) and (4.20)-(4.23) we find that we have an expansion
for Det$(n)$ in terms of manifestly gauge invariant quantities:  Traces of
line integrals around loops, with some insertions of $F_{\mu \nu}$'s in the
loops.

We turn to the denominator determinant Det$(b)$ from (4.8).  We first make
some definitions
\begin{eqnarray}
b_{\alpha \beta}(s_1,\;s_2) &=& \Big< \phi^{s_1}_\alpha, \;
\psi^{-s_2}_\beta \Big> \\
B_{\alpha \beta}(s_1,\;s_2) &=& \Big< \phi^{s_1}_\alpha, \;
\phi^{-s_2}_\beta \Big>
\end{eqnarray}
and note that the matrix $b$ from (4.10) is given as
\begin{equation}
b_{\alpha \beta} \ = \ b_{\alpha \beta}(1,1)
\end{equation}
We note from the fundamental theorem of calculus
\begin{eqnarray}
\left|{\rm Det}\Big(b_{\alpha \beta}(1,1) \Big)\right|^2 &=& \left|{\rm Det}
\Big(b_{\alpha \beta}(0,0) \Big)\right|^2 \cdot e^{\int^1_0 d\lambda {d \over
d\lambda} \; ln \; \big|{\rm Det} \big(b_{\alpha \beta}(\lambda,\lambda)
\big)\big|^2} \\
&=& {\rm Det} \Big(B_{\alpha \beta}(0,0) \Big) \cdot e^{2 {\rm Re} \int^1_0
d\lambda {d \over {d \lambda}} \; P \big(b(\lambda,\lambda) \big)}
\end{eqnarray}
where we have defined $P$ by
\begin{equation}
{\rm Det}(A) = e^{P(A)} = e^{Tr \left[ (A-\delta) - {1 \over 2}
(A-\delta)^2 \cdots \right] }.
\end{equation}
We now study the derivative in (4.28)
\begin{equation}
{d \over d\lambda} \; P\big(b(\lambda,\lambda)\big) =
{d \over d\ell} \; P\big(b(\ell,\lambda)\big) \Big|_{\ell = \lambda}   +
{d \over dr} \; P\big(b(\lambda , r)\big) \Big|_{r=\lambda}
\end{equation}

The terms on the right side can be reexpressed using the following
surprising insights:
\begin{description}
\item [1)]
\begin{equation}
{d \over d\ell} \; P\big(b(\ell,\lambda )\big) \Big|_{\ell=\lambda}
= {d \over d\ell} \; P\big(B(\ell,\lambda )\big) \Big|_{\ell=\lambda}.
\end{equation}
This follows from the determinant product formula:
\begin{equation}
P(AB) \ = \ P(A) + P(B)
\end{equation}
and rewriting
\begin{equation}
B(\ell,r) = b(\ell,r) b^*(-r,-r).
\end{equation}
\item [2)]
The last derivative term in (4.30) is independent of the gauge field
$A_\mu(x)$.  We see this by a similar argument to that in 1)
using instead of (4.33) the following
\begin{equation}
b(\lambda, r) = b(\lambda,\lambda)B^0(\lambda,r)
\end{equation}
with
\begin{equation}
B^0_{\alpha \beta}(s_1,s_2) = \Big< \psi^{s_1}_\alpha, \; \psi^{-s_1}_\beta
\Big> .
\end{equation}
\end{description}
In the arguments of 1) and 2) we have used the completeness relation, for
each $s$
\begin{equation}
\sum_\alpha \left| \psi^s_\alpha \  >< \  \psi^{-s}_\alpha \right| \  =
\delta .
\end{equation}

We then rewrite (4.28)
\begin{equation}
 |{\rm Det} \Big(b(1,1) \Big)|^2 = e^{P(B(0,0))} \cdot e^{2 {\rm Re} \int^1_0
d\lambda {d \over {d \ell}} \; P \big(B(\ell,\lambda)
\big)\big|_{\ell=\lambda}}.
\end{equation}
The advantage of the form of equation (4.37) to study the denominator of
(4.8), is that $P(B)$ is given in terms of obviously gauge invariant
quantities.  We turn to seeing this.

We first note, for $B = B(\ell,-r)$,
\begin{equation}
{\rm Det}(B) = e^{P(B)} = e^{Tr\big((B-\delta) - {1 \over 2}(B-\delta)^2
\cdots \big)}.
\end{equation}
We study Tr$\big((B-\delta)^s \big)$.  We let $c$ be a sequence of $s$
wavelets
\begin{equation}
c = \left( \beta_1, \beta_2, \cdots , \beta_s \right)
\end{equation}
and then express the trace
\begin{equation}
Tr \Big((B-\delta)^s\Big)=\sum_c Tr \Big( \Big<
\phi^\ell_{\beta_1},\phi^r_{\beta_2}\Big>  \cdots \Big< \phi^\ell_{\beta_s},
\phi^r_{\beta_1}\Big> \Big)
\end{equation}
where on the right side the trace is in group space only.  We define
\begin{equation}
f_c(x_1,\cdots,x_s) = \Big( \psi^\ell_{\beta_1}(x_1) \psi^r_{\beta_2}(x_1)
\Big) \Big( \psi^\ell_{\beta_2}(x_2) \psi^r_{\beta_3}(x_2) \Big) \cdots
\Big( \psi^\ell_{\beta_s}(x_s) \psi^r_{\beta_1}(x_s) \Big)
\end{equation}
and then have
\begin{equation}
Tr \Big((B-\delta)^s - (B^0-\delta)^s \Big)  = \sum_c \int dx_1 \cdots \int
dx_s \ f_c(x_1, \cdots , x_s) \cdot Tr [L - 1]
\end{equation}
with $L$ the closed loop line integral
\begin{equation}
L = L(c,x_1, \cdots , x_s)=u(1,x_1) u(x_1,2)u(2,x_2)u(x_2,3)\cdots u(x_s,1)
\end{equation}
and $i$ stands for $\gamma_{\beta_i}$.

Thus the denominator determinant is
expressed in terms of traces of closed loop line integrals, a manifestly
gauge invariant expression.
\vfill\eject

\chapter{Determinant of a Gauged Fermion Field in an External Gauge Field}

We will not repeat details that are exactly parallel to the boson case of
the last chapter, and instead highlight new features.

We let
\begin{equation}
{\cal L} = \gamma^\mu(\partial_\mu + A_\mu) + m
\end{equation}
and our object of study is
\begin{equation}
M = M(A) = {\rm Det}({\cal L})
\end{equation}
We, without loss of generality, set
\begin{equation}
m \ \ge \  0.
\end{equation}
In the $m=0$ case we do not consider $\gamma_5$ couplings, i.e. setting
\begin{equation}
A_\mu = A^1_\mu + A^2_\mu \, \gamma^5
\end{equation}
with $A^i_\mu$ in the gauge group Lie algebra.  We would like to learn the
right way to handle such chiral couplings in the present formalism.

We set
\begin{equation}
{\cal L}_0 = \gamma^\mu \partial_\mu + m.
\end{equation}
We note that ${\cal L}_0$ is not self-adjoint, but it is normal; it
commutes with its adjoint.  What we actually use is that
\begin{equation}
{\cal L}_0 = \alpha + i \beta
\end{equation}
with $\alpha$ a number $\ge 0$, and $\beta$ a self-adjoint operator.

Parallel to (2.5) and (4.4) we define (suppressing spinor indices):
\begin{eqnarray}
f^s_\alpha &=& \big({\cal L}_0\big)^{- {s \over 2}} \ \psi_\alpha \\
F^s_\alpha &=& u(x,\gamma_\alpha) f^s_\alpha \\
\hat{f}^s_\alpha &=& \big(({\cal L}_0)^{- {s \over 2}}\big)^* \ \psi_\alpha \\
\hat{F}^s_\alpha &=& u(x,\gamma_\alpha)\; \hat{f}^s_\alpha.
\end{eqnarray}
Fractional powers of ${\cal L}_0$ are taken using the fractional calculus
with a branch cut along the negative real axis, working with the Principal
branch.

In inner products functions on the left are always caretted. Equations
(4.36), (4.24), (4.25), (4.8), (4.9) respectively are replaced by:
\begin{eqnarray}
\sum_\alpha \left|f^s_\alpha \ >< \ \hat{f}^{-s}_\alpha \right| &=& \delta \\
b_{\alpha \beta}(s_1,s_2) &=& \Big< \hat{F}^{s_1}_\alpha, f^{-s_2}_\beta
\Big>, d_{\alpha \beta}(s_1, s_2) = \Big< \hat{f}^{s_1}_\alpha,
F^{-s_2}_\beta \Big> \\
B_{\alpha \beta}(s_1,s_2) &=& \Big< \hat{F}^{s_1}_\alpha,\; F^{-s_2}_\beta
\Big> \\
{\rm Det}({\cal L}) &=& \ {\rm Det}(n_{\alpha \beta})/ \Big({\rm
Det}(b(1,1)){\rm Det}(d(1,1))\Big) \\
n_{\alpha \beta} &=& \Big< \hat{F}^1_\alpha, \; {\cal L} \ F^1_\beta \Big>
\end{eqnarray}
as the most important examples of transcriptions.  We must work with
Det$(bd)$ instead of $|{\rm Det}(b)|^2$, but the development is
straightforward with the same ideas as in the last chapter.  Again
Det$({\cal L})$ will be expressed in terms of explicitly gauge invariant
quantities.

\vfill\eject

\centerline{REFERENCES}
\begin{description}
\item [[1]] J. Glimm, A. Jaffe, ``Positivity of the $\phi^4_3$ Hamiltonian'',
{\it Fort. Phys.} {\bf 21}, 327-376 (1973).
\item [[2]] G.A. Battle, P. Federbush, ``A Phase Cell Cluster Expansion for
Euclidean Field Theories'', {\it Annals of Phys.} {\bf 142}, 95-139 (1982).

\item [\ \ \ \ ] G.A. Battle, P. Federbush, ``A Phase Cell Cluster Expansion
for a Hierarchical $\phi^4_3$ Model'', {\it Commun. Math. Phys.} {\bf 88},
263-293 (1983).
\item [\ \ \ \ ] G.A. Battle, P. Federbush ``Ondelettes and Phase Cell
Cluster Expansions, A Vindication'', {\it Commun. Math. Phys.} {\bf 109},
417-419 (1987).
\item [[3]] P. Federbush, ``Quantum Field Theory in Ninety Minutes'', {\it
Bull. A.M.S.} {\bf 17}, 93-103 (1987).
\item [[4]] P. Federbush, ``On the Quantum Yang-Mills Field Theory'', in
{\it Mathematical Quantum Field Theory and Related Topics}, Proceedings of
the 1987 Montreal Conference held Sept. 1-5, 1987 (CMS Conference
Proceedings, Vol. 9).
\item [\ \ \ \ ] P. Federbush, ``A Phase Cell Approch to Yang-Mills Theory, IV,
The Choice of Variables'', {\it Commun. Math. Phys.} {\bf 114}, 317-343
(1988).
\item [[5]] J. Magnen, V. Rivasseau, R. Seneor, ``Construction of $ym_4$ with
an infra-red cutoff'', {\it Commun. Math. Phys.} {\bf 155}, 325-383 (1993).
\item [[6]]  T. Balaban, A. Jaffe, ``Constructive Gauge Theory'', in G. Velo
and A.S. Wightman, editors, {\it Fundamental Problems of Gauge Field
Theory. Erice 1985}, 207-263, New York, 1985.  Erice Summer School, Plenum
Press.
\item [[7]] L. Rosen, ``Perturbative Renormalization'' in {\it Canadian
Mathematical Society, Conference Proceedings}, Montreal, Volume 9, (1988).
\item [\ \ \ \ ] J.Feldman, T. Hurd, L. Rosen, and J. Wright, ``QED: A Proof of
Renormalizability'', {\it Springer Lecture Notes in Physics}, Vol. 312,
Springer Verlag (1988).
\item [\ \ \ \ ] G. Gallavotti, ``Renormalization Theory and Ultraviolet
Stability for Scalar Fields Via Renormalization Group Methods'',
{\it Rev. Mod. Phys.} {\bf 57} (1985) 471-562.
\item [\ \ ] G. Gallavotti and F. Nicolo, ``Renormalization Theory in Four
Dimensinal Scalar Fields'', {\it Commun. Math. Phys.} {\bf
100} (545-590) (1985) and {\bf 101} 247-282 (1986).
\item [[8]] G.`t Hooft, M. Veltman, ``Regularization and Renormalization of
Gauge Fields'', {\it Nucl. Phys.} {\bf B44}, 189-213 (1972).
\item [[9]] P. Federbush, ``A Hierarchical Small Field Non-Linear Sigma
Model'', in Proceedings of the Conference on Probability Models in
Mathematical Physics, Colorado Springs, May 1990, editors, G.J. Morrow and
W-S Yang, {\it World Scientific}.
\item [[10]] Y. Meyer, ``Principe d'incertitude, bases hilbertiennes et
alg\`{e}bres d' op\'{e}rateurs'', {\it Seminaire Bourbaki} {\bf 38}, 662
(1985-86).
\item [[11]]  G.A. Battle, P. Federbush, P. Uhlig, ``Wavelets for Quantum
Gravity and Divergence-Free Wavelets'', {\it Appl. and Comp. Harmonic
Anal.} {\bf 1}, 295-297 (1994).
\end{description}

\end{document}